%% file: main.tex
\documentclass[conference]{IEEEtran}
\usepackage{cite}

\usepackage{amsmath,amssymb,amsfonts}
\usepackage{algorithmic}
\usepackage{graphicx}
\usepackage{textcomp}
\usepackage{xcolor}
\def\BibTeX{{\rm B\kern-.05em{\sc i\kern-.025em b}\kern-.08em
    T\kern-.1667em\lower.7ex\hbox{E}\kern-.125emX}}

\usepackage[ruled,vlined,lined,commentsnumbered]{algorithm2e}
\usepackage{balance}
\usepackage{csvsimple}
\usepackage{longtable}
\usepackage{array, tabularx}
\usepackage{booktabs}
\usepackage[skip=1pt,font=small,labelfont=bf,textfont=bf]{caption}
\usepackage{url}
\usepackage{lscape}
\usepackage{rotating}
\usepackage{tikz}
\usepackage[normalem]{ulem}
\usepackage{enumitem}
 \usepackage{array,multirow,graphicx}
 \usepackage{float}
\usepackage{amsmath}
\usepackage{amssymb}
\usepackage{fancybox}
\usepackage{caption}
\usepackage{subcaption}
\usepackage{hyperref}
\usepackage{colortbl}
\usepackage[most]{tcolorbox}

\usepackage{threeparttable}

\usepackage{marvosym}
\usepackage{textcomp}
\usepackage{arydshln}
\usepackage{tikz}

\setlist{noitemsep} 

\def\BibTeX{{\rm B\kern-.05em{\sc i\kern-.025em b}\kern-.08em
    T\kern-.1667em\lower.7ex\hbox{E}\kern-.125emX}}

\newcommand{\find}[1]{\begin{tcolorbox}[leftrule=1mm,rightrule=1mm,toprule=0mm,bottomrule=0mm,left=2pt,right=2pt,top=0pt,bottom=0pt]
#1
\end{tcolorbox}
}

\definecolor{codegreen}{rgb}{0,0.6,0}
\definecolor{codegray}{rgb}{0.5,0.5,0.5}
\definecolor{codepurple}{rgb}{0.58,0,0.82}
\definecolor{backcolour}{rgb}{0.95,0.95,0.92}
 
\definecolor{DarkOrange}{rgb}{0.8,0.3,0.0} 
\definecolor{yellow}{RGB}{255,255,153}
\definecolor{grey}{RGB}{224,224,224}

\newboolean{showcomments}
 \setboolean{showcomments}{true}
\ifthenelse{\boolean{showcomments}}
 { \newcommand{\mynote}[2]{
      \fbox{\bfseries\sffamily\scriptsize#1}
        {\small$\blacktriangleright$\textsf{\emph{#2}}$\blacktriangleleft$}}}
        { \newcommand{\mynote}[2]{}}

\newcommand{\datasetName}{\texttt{SBR}\xspace}

\AtBeginDocument{%
  \providecommand\BibTeX{{%
    \normalfont B\kern-0.5em{\scshape i\kern-0.25em b}\kern-0.8em\TeX}}}

\begin{document}

\title{Early Detection of Security-Relevant Bug Reports
using Machine Learning: How Far Are We?}


\author{Arthur D. Sawadogo*, Quentin Guimard*, Tegawendé F. Bissyandé†, \\
Abdoul Kader Kabore†, Jacques Klein†, Naouel Moha*\\
*Université du Québec à Montréal\\
†University of Luxembourg\\
}

\maketitle








\begin{abstract}
 \input{abstract.tex}
\end{abstract}


\input{introduction.tex}

\input{motivation.tex}
\input{state.tex}

\input{dataset.tex}

\input{engineering.tex}

\input{related.tex}

\input{conclusion.tex}

\bibliographystyle{IEEEtran}
\bibliography{reff}
\end{document}

%% file: abstract.tex
Bug reports are common artefacts in software development. They serve as the main channel for users to communicate to developers information about the issues that they encounter when using released versions of software programs. In the descriptions of issues, however, a user may, intentionally or not, expose a vulnerability. In a typical maintenance scenario, such security-relevant bug reports are prioritised by the development team when preparing corrective patches. Nevertheless, when security relevance is not immediately expressed (e.g., via a tag) or rapidly identified by triaging teams, the open security-relevant bug report can become a critical leak of sensitive information that attackers can leverage to perform zero-day attacks. To support practitioners in triaging bug reports, the research community has proposed a number of approaches for the detection of security-relevant bug reports. In recent years, approaches in this respect based on machine learning have been reported with promising performance.  Our work focuses on such approaches, and revisits their building blocks to provide a comprehensive view on the current achievements. To that end, we built a large experimental dataset and performed extensive experiments with variations in feature sets and learning algorithms. Eventually, our study highlights different approach configurations that yield best performing classifiers.

%% file: introduction.tex
\section{Introduction}
\label{sec:introduction}

Bug tracking systems are commonplace in both open source software development and large industrial software projects.
Their consistent use as part of software project management is regarded as one of the {\em hallmark of a good software development team}~\cite{spolsky2010painless}. 
In industrial settings, development teams can setup bug tracking systems to keep a centralized view of all bug fixes (and potentially improvements) brought to the code base. 
In open source communities, they allow end-users to report bugs directly to the development teams.

A bug tracking system mainly includes a database of bug reports, which are natural language descriptions of erroneous behavior cases that are encountered when running the software. 
A given record (i.e., a bug report's details in the database) may include the identity of the person who reported it, the time it was reported, its severity and other tags that can be leveraged for facilitating triaging. 
When the {\em severity} of the bug is major (e.g., a blocking bug~\cite{li2019watch}) or when the bug is relevant to {\em security}, development teams urge to address the bug report. 
Sometimes, however, the person who reported a problem with respect to end-user requirements, may be unaware that the deviated behavior is actually also offering a vulnerability window for attackers to misuse the software. 
Thus, unintentionally, such {\em security-relevant} bug reports can publicly disclose exploitable vulnerabilities.  
Until the triaging team can notice the security risk (and therefore take actions to hide\cite{mostafaautomatic,Mostafa2019} the bug report until a fix is found), any malicious actor can leverage the divulged input and perform zero-day attacks.

Security-relevant bug reports can be categorized as Hidden Impact Bugs~\cite{wijayasekara2014vulnerability}: the bugs are often only identified as vulnerabilities long after they have been made public. 
Identifying security-relevant bug reports is thus an important challenge for bug triagers. Automating this task will consequently improve bug report management where vulnerabilities are assigned immediately to the appropriate developers and are patched in priority. An early attempt to automating the detection of security-relevant bugs was performed by Wijayasekara et al. \cite{wijayasekara2012mining,wijayasekara2014vulnerability} focusing on Linux kernel CVEs\footnote{Common Vulnerability Exposure} that are associated to bug reports. Their classifier was based on a basic ``bag of words'' approach in combination with Naive Bayes, Naive Bayes Multinomial, and Decision Tree classifiers.
Behl et al.~\cite{Behl2014} then used the Term Frequency-Inverse Document Frequency (TF-IDF) along with a ``vector space model'' and showed that it provides better performance than the Naive Bayes algorithm. More recently, Peters et al.~\cite{Peters2019} proposed a framework
called FARSEC, which integrated filtering and ranking for
security bug report prediction: prior to creating the prediction
models, FARSEC identifies and removes non-security bug
reports with security related keywords. This filtering step is mainly aimed at decreasing the false positive predictions. 
Unlike previous related works, which all used prediction models based on machine learning, Goseva and Tyo~\cite{Goseva-Popstojanova2018} also proposed to investigate unsupervised learning algorithms and assessed the importance of the size of the dataset. 
More recently, Shu et al.~\cite{shu2019better} studied hyperparameter tuning and data pre-processing to improve the performance of security-relevant bug report classification.

Overall, it appears that machine learning has become a central technique for identifying security-relevant bug reports. The state of the art in this respect remains however unclear about the limitations and promises of the different learning choices. Thus, we propose in this paper to present extensive experiments on features and algorithms for establishing the achievements that can be reached. Our paper makes the following contributions to the field:

\begin{itemize}
	\item We discuss motivational cases for the research direction on detecting security-relevant bug reports, and provide an overview of the state of research.
	\item We prepare and offer to the community a large dataset of security-relevant bug reports to enable comprehensive investigations around this topic.
	\item We performed extensive experiments on learning to identify security bug reports, and enumerate insights about the variety of algorithms, features and parameters.
\end{itemize}

{\bf Paper Organization.} The remainder of this paper is organized as follows. Section~\ref{subsec:motivation} highlights the importance of security-relevant bug report identification through motivating cases. Section~\ref{sec:survey} discusses the state of the art approaches for identifying security-relevant bug reports. We then overview the construction of a large experimental dataset of security-relevant bug reports in Section
~\ref{sec:datasets}. We present our experimental results in Section~\ref{sec:experiments} and discuss threats to validity as well as similar works in Section~\ref{sec:discussion}  before concluding the paper in Section~\ref{sec:conclusion}. 


%% file: motivation.tex
\section{On the Importance of Identifying Security-Relevant Bug Reports}
\label{subsec:motivation}
Although the detection of security bug reports is increasingly studied, the literature does not include data or motivational cases to support the importance of the topic. In this section, we attempt to fill this gap through (1) an empirical study to highlight how security bug reports, when identified, are processed in priority, (2) a highlight of cases where bugs are reported by users who are not aware of the security-relevance and the potential for attacks.

\subsection{Priority patching by development teams}
Our first investigation is related to the priority that is given to security bug reports. 
We aim to answer the following question: ``{\em Are security-relevant bug reports responded with faster patches than other bug reports?}'' When a bug report is submitted, developers generally attempt to address it and eventually propose a patch to fix the relevant bug. Such a patch can however take more or less time to be prepared. Nevertheless, we expect that the process will be relatively fast if the bug report is about a vulnerability that could be exploited by attackers.

To answer our research question, we consider bug report samples from our ground truth dataset (cf. Section~\ref{sec:datasets} for more details on dataset collection). Overall, we focused on RedHat bug reports and randomly sampled 800 in each category (i.e., security bug reports and non-security bug reports). We then compute the delays between the creation date of a bug report, and its closing date. Identify the exact fix date of a bug report is not obvious, we then rely on opening and closing dates in order to have an accurate view. While these dates do not confirm that bug reports were closed at the same time as the reported bug was fixed, they give us an idea of how important it is to identify security-relevant bug reports early on.

\begin{figure}[!h]
\centering
\includegraphics[width=1.00\linewidth]{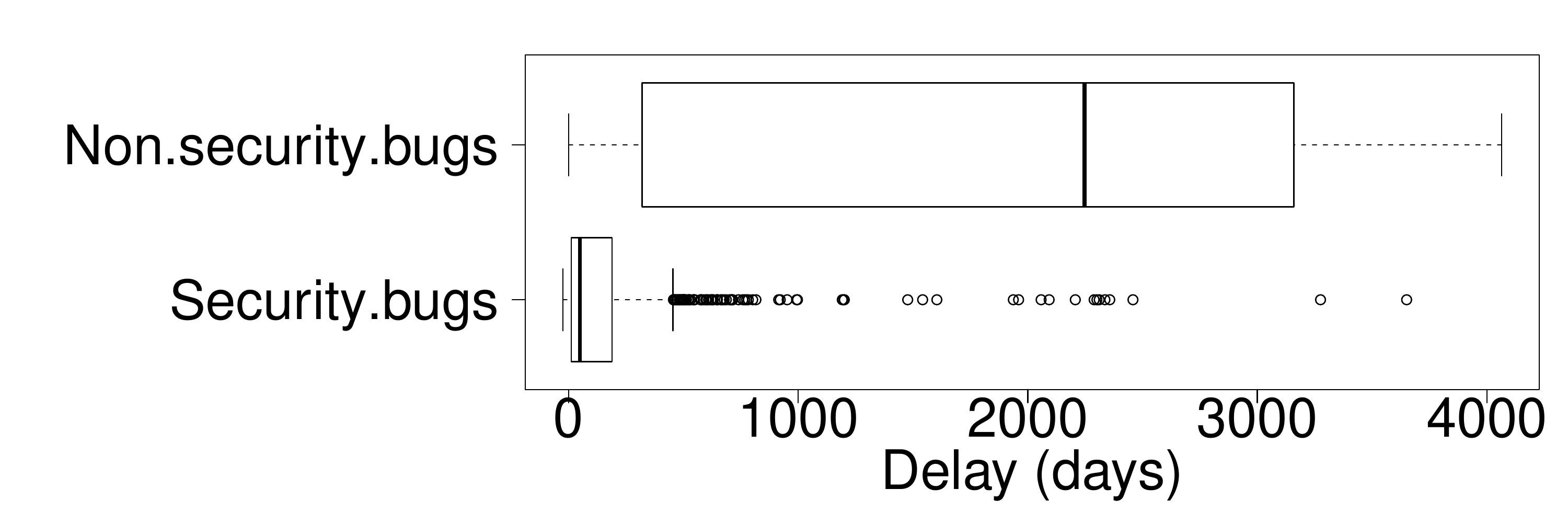}
\caption{ Delays between creation date and closing date of security bug reports vs other bugs reports}

\label{fig:delays-patching}
\end{figure}

Figure~\ref{fig:delays-patching} presents the distribution of time-to-fix delays for the two categories of bug reports. 
The median values (2248.0 days and 49.5 days) as well as the ranges of the quartiles confirm that security bug reports are addressed significantly faster than other bug reports. Given that the dataset was constructed based on the security tag used in the bug tracking system, this finding suggests that it is important to appropriately tag bug reports in order to accelerate their processing. 
An automatic approach to identify bug reports that are related to security is thus desirable to alleviate triaging effort.

\find{If proper notice is given, bugs triagers are likely to prioritize addressing security-relevant bug reports}
\subsection{Unintentional disclosure of vulnerabilities}


In a typical bug reporting scenario, a user is affected by an erroneous software behavior and writes a description of the bug for the development team.
When such a description refers to information such as CVE details or exploitation cases, the triaging team can immediately tag the bug report as being security-relevant. 
Such a bug report is generally handled with high priority by maintainers, and a patch is rapidly constructed. 
Unfortunately, there are reporting scenarios where the bug report, although being security-relevant, is not tagged as such due to the lack of explicit details that would inform triagers immediately. 
In such cases, a software vulnerability is potentially exposed publicly. 
Security relevance of the bug report can then be later pointed out by other contributors. 
In the mean time, however, the software is exposed to attacks, because the delay between the bug's reporting and its prioritization as security-relevant could be important. 
Attackers (like all other parties interested in contributing to the software development in open source settings), are handed critical information for exploiting software vulnerabilities. These attackers even have the opportunity to quickly propose a patch that satisfies the bug reporter (in terms of user-expected plain functionality) but better ensures that the vulnerability remains hidden for their future exploitations.

We look for evidence about these conjectures by carefully investigating 100 security bug reports randomly sampled in our collected dataset. 
We mainly look for bug reports that were not immediately flagged as security-relevant but were identified as such only later by other contributors. 
The existence of such bug reports justify the need for automated techniques for early detection of security relevance. In the following, we present two out of several cases that we have found.

Figure~\ref{fig:sbr1} shows an excerpt of bug report N$^o$ 319004 in the Bug tracking system of Mozilla. Obviously, the bug reporter did not seem to know that the bug was related to security aspect: she only reports a bothersome development issue and is looking for a solution. During discussions, four days later, a seventh comment raises concerns about potential exploitations by attackers. A CVE (CVE-2005-4134) will be filed only the following day.

\begin{figure}[hhh]
	{\parbox{\linewidth}{
\begin{lstlisting}[linewidth={\linewidth},frame=tb,basicstyle=\scriptsize\ttfamily]

User2 (:shaver -- probably not reading bugmail closely)

Description 
From the URL:

Basically firefox logs all kinda of URL data in it's 
history.dat file, this little script will set a really 
large topic and Firefox will then save that topic into
it's history.dat.. The next time that firefox is opened, 
it will instantly crash due to a buffer overflow -- this 
will happen everytime until you manually delete the 
history.dat file -- which most users won't figure out.

We're probably blowing Mork's little mind, and it'll 
be easier to defend in the history code than in Mork.

People are reporting varying results in #developers.

       -----------------------------------------------

elfguy:
Comment 7 
This is being reported as 'remote exploit' in the news:

http://digg.com/security/Kill_Firefox_1.5_with_remote_exploit

While it's not a security issue it should be fixed fast as 
malicious users may start using it now that it's known.
\end{lstlisting}
	}}%
	\caption{Example of a security Bug report from Buzilla-mozilla}
	\label{fig:sbr1}
\end{figure}

Figure~\ref{fig:sbr2} presents bug report APLO-366 in the bug tracking system of JIRA. Similar to the previous case, the bug is reported without any tag related to security. Instead, the linked patch is commented as referring to a CVE tag (CVE-2014-3579) being addressed. Nevertheless, the patch itself does not mention any security matter in its description, leading to a silent security patching, with all its consequences.

\begin{figure}[hhh]
	{\parbox{\linewidth}{
\begin{lstlisting}[linewidth={\linewidth},frame=tb,basicstyle=\scriptsize\ttfamily]
XPath selector - make xml parser features configurable
	Details
	Type: Bug
	Status:RESOLVED
	Priority: Major
	Resolution: Fixed
	Affects Version/s:
	1.7
	Fix Version/s:
	1.7.1, 1.8
	Component/s: None
	-Created: 26/Aug/14 13:05

Description
	We need to be able to set xml parser features 
	configurable using system properties.

    ----------------------------------

Comments
User1 added a comment 26/Aug/14 13:29

Fixed with http://git-wip-us.apache.org/repos/asf/
activemq-apollo/ commit/e5647554  with some sensible 
defaults introduced. Features can be set by using system 
properties starting with "org.apache.activemq.apollo.
documentBuilderFactory.feature:" prefix, like
-Dorg.apache.activemq.apollo.documentBuilderFactory.feature:
http://xml.org/sax/features/external-general-entities=true 

User1 added a comment - 06/Feb/15 10:44

This issue fixes CVE-2014-3579. Take a look at 
http://activemq.apache.org/security-advisories.html 
for more details.

\end{lstlisting}
	}}%
	\caption{Example of a security Bug report from Jira-apache}
	\label{fig:sbr2}
\end{figure}

\find{Security-relevant bug reports could expose the software to more attacks if they are not tagged as such earlier.}


The presented examples further motivate the necessity to automate the detection of security-relevant bug reports to enable appropriate management. 

%% file: state.tex
\section{A  Review of the State of the Art}
\label{sec:survey}
In this section, we present the state of the research related to the detection of security relevant bug reports. Since security relevant bug reports are referred to with different terminology in the literature, we propose to rely on a light-weight snowballing approach to identify relevant papers. Thus, we start with two recent papers in the literature, namely in 2019 by Shu et al.~\cite{shu2019better} and Pereira et al.~\cite{Pereira2019}, and check their related work sections. Then, we iteratively look into the related work and references of the identified relevant papers. Eventually, we found that most of the related work rely on machine learning (ML) based approaches. 
In a first sub-section, we summarize approaches grouped based on the type of learning (i.e., supervised, semi-supervised, unsupervised). 
Then, we list the main feature extraction methods used to feed the machine learning algorithms. Finally, we quickly enumerate the various validation approaches used to assess the performance of the ML-based state of the art approaches. 
Overall, Table~\ref{tab:fps} provides a summary of information collected during our review of the state of the art.

\begin{table*}[!h]
	\centering
	\scriptsize
	\caption{Summary details of state of the art approaches for security bug reports identification.}
	\resizebox{1\linewidth}{!}{
	\begin{threeparttable}
		\begin{tabular}{lp{2cm}p{2cm}p{2cm}p{2cm}p{2cm}p{1cm}p{1cm}}
		 {\bf \normalsize Approach} & {\bf \normalsize Features}& {\bf\normalsize Type of Approach} & {\bf\normalsize Algorithm}& {\bf\normalsize Evaluation} &{\bf\normalsize Dataset} &{\bf\normalsize \# SBR$^\ast$}&{\bf\normalsize \# All BR$^\ast$}\\
		\toprule
		\normalsize Gegick et al.~\cite{Gegick2010}&\normalsize hypothetical term-by-document matrix &\normalsize Supervised &\normalsize Statistical model&\normalsize Automatic &\normalsize Cisco software system&&\\
		\midrule
		\normalsize Wijayasekara et al~\cite{wijayasekara2014vulnerability} &\normalsize Most frequent words&\normalsize Supervised &\normalsize Naives bayes, Decision Trees, etc.&\normalsize Automatic &\normalsize linux&&\normalsize6000\\
		\midrule
		\normalsize Behl et al.~\cite{Behl2014}&\normalsize  TF-IDF&\normalsize supervised &\normalsize Naives bayes&\normalsize Automatic &\normalsize Bugzilla&&\normalsize 10000\\
		\midrule
		\normalsize Zou et al.~\cite{Zou2018} &\normalsize  Meta features and textual features &\normalsize Supervised &\normalsize SVM&\normalsize Automatic &\normalsize Firefox, Thunderbird, Seamonkey &\normalsize3346&\normalsize 23262\\
		\midrule
		\normalsize Goseva-Popstojanova et al.~\cite{Goseva-Popstojanova2018}&\normalsize  TF,TF-IDF,BF&\normalsize Supervised, unsupervised &\normalsize SVM, KNN, etc.&\normalsize Automatic &\normalsize NASA dataset&\normalsize 664&\normalsize3399\\
		\midrule
		\normalsize Mostafa et al.~\cite{Mostafa2019}&\normalsize TF-IDF, descrimitive weight &\normalsize Supervised, semi-supervised &\normalsize SAIS,SVM, etc. &\normalsize Automatic &\normalsize Rehdat, Mozilla&\normalsize 1368&\normalsize 16168\\
		\midrule
		\normalsize Das et al.~\cite{Das2019}&\normalsize TF-IDF &\normalsize Supervised &\normalsize Naive Bayes Multinomial Model&\normalsize Automatic &\normalsize Ambari, Camel, Derby, Wicket&\normalsize 159&\normalsize 4000\\
		\midrule
		\normalsize Pereira et al.~\cite{Pereira2019}&\normalsize TF-IDF &\normalsize Supervised &\normalsize Adaboost, Naives bayes,etc.&\normalsize Automatic &\normalsize Microsoft&\normalsize 552073&\normalsize 1073149\\
		\midrule
	\normalsize	Peters et al.~\cite{Peters2019}& \normalsize TF-IDF, score&\normalsize Supervised &\normalsize logistic regression, Naives bayes, etc.&\normalsize Automatic &\normalsize Chromium, Wicket, Ambari, Camel, Derby&\normalsize 351&\normalsize 45940\\
		\bottomrule
		\end{tabular}
	\normalsize	$^\ast$: the "\# SBR" column indicates the number of security relevant bug reports used in an approach.  The "\# All BR" column indicates the number of security and non-security bug reports used in an approach.
	\end{threeparttable}
	}
	\label{tab:fps}
\end{table*}


\subsection{Overview of ML-based Approaches}

\paragraph{\textbf{Supervised learning approaches}}
Supervised learning is a classical prediction problem in which there are two classes and labeled data from both classes (positive class as the targeted class and negative class) are available. The chosen machine learning algorithm is trained on one part of the data and predict on the other part. The main problem of this approach is that it is often difficult for a given problem to have labeled data from both classes. It could happen in some cases that only one class is labeled.
Regarding the case of security bug report detection, the prediction problem resides in differentiating security relevant bug reports from non-security relevant bug reports. 

Supervised learning is the most widespread approach leveraged in the literature for security bug report identification. 
Wijayasekara et al.~\cite{wijayasekara2014vulnerability} have presented one of the seminal work on detecting security bug reports using machine learning. They rely on text mining to extract syntactical information bug reports and compress before generating feature vectors that are fed to Naive Bayes classifiers. The proposed methodology was tested on Linux vulnerabilities that were discovered in the time period from 2006 to 2011, where they managed to correctly identify up to 88\% of vulnerabilities. 
Such an approach improves over the work of Gegick et al.~\cite{Gegick2010} who used a term-by-document frequency matrix from words in the natural language descriptions of bug reports to train a statistical model. Similarly, Behl et al.~\cite{Behl2014}, later compared the use of term frequency-inverse document frequency (TF-IDF) against a probabilistic learning approach like Naives Bayes.

Zou et al.~\cite{Zou2018} proposed to use a combination of text-mining features and meta-data (e.g., time, severity, and priority) for improving identification of security bugs reports. They trained a supervised approach (SVM) with Radial Basis Function( RBF) and have managed to improve previous work by over 20 percentage points. 

More recently, Das et al.~\cite{Das2019} proposed an approach based on class imbalance sampling and TF-IDF vectors to improve security-relevant bug report detection using Naive Bayes Multinomial classification.
Pereira et al.~\cite{Pereira2019} also presented preliminary results, which show that basic TF-IDF on just the title of bug reports may be enough to provide accurate results in detecting security bug reports.


Following up on this state of the art investigations, Peters et al.~\cite{Peters2019}  proposed FARSEC, a framework for filtering and ranking bug reports for reducing the presence of security-related keywords and improving text-based prediction models for security bugs fixes. 
Their framework is presented as improving the use of text-mining approaches (TF-IDF, DTF, IDF, etc.) when considering feature extraction for machine learning.

\paragraph{\textbf{Semi-supervised learning approaches}}
Semi-supervised learning represents the case where only one class is labeled and correctly identified. Several algorithms perform semi-supervised learning for identifying security bug reports with good prediction results. However, This approach, although very interesting, got bad performances when the positive set is not enough representative for training.

Mostafa et al.~\cite{Mostafa2019} recently presented an evolutive and realistic approach for the identification of security bug reports which considers the evolution of security vocabulary on NVD database and realistic constraints like small training set for security bugs reports prediction. They train a semi-supervised model to predict security bugs reports that will be used to augment the training set of a fully supervised learning model.

\paragraph{\textbf{Unsupervised learning approaches}}
The main difference between the previous approaches (supervised and semi-supervised) and the unsupervised approaches is the fact that in the unsupervised case the data are not labeled. The algorithm must then train and predict from unlabeled examples. Clustering algorithms are frequently used to group data with similar characteristics from a classification point of view. 

In a recent study  Goseva-Popstojanova et al.~\cite{Goseva-Popstojanova2018} assess the impact of algorithms and features in the detection of security bug reports. They offer one the rare works that explore unsupervised learning algorithms to regroup security bug reports. Their preliminary results however shows that supervised learning algorithms offer higher performance metrics.

\paragraph{\textbf{Deep learning approaches}} Since traditional machine learning requires manual feature engineering, its performance is highly dependent on the quality of the features identified by researchers or practitioners. Instead, many recent works turn to deep learning for learning the the best representation and then applying on top the automatically-extracted feature vectors a traditional algorithm. 

In a recent work, Kukkar et al.~\cite{kukkar2019novel} have proposed a deep learning model for multiclass severity classification called Bug
Severity classification using a Convolutional Neural Network and
Random forest with Boosting. They show that the Convolutional Neural Network is able to extract the important
feature patterns of respective severity classes. Although they did not apply their work on the problem of security classification, to the best of our knowledge, it is one of the most related approach that leverages deep learning.



 \subsection{Feature extraction}
 Since the large majority of the state of the art rely on traditional machine learning, we investigate the leaning pipeline for reproducing different approaches. Key steps in a machine learning pipeline are feature engineering (i.e. what are the key ingredients that make machine learning algorithms work) and feature extraction (the process to extract and compute the features from the input samples).  
In the case of the security bug reports detection task,
data collected (i.e., the bug reports) might not be directly readable by algorithms in their default nature. 
The feature extraction step is then required. 
We consider in this study many feature extraction approaches used in the state of art for extracting representative information in security bug reports detection. We can summarize the different approaches to feature extraction by:



\begin{itemize}
	\item \textbf{Frequencies of words in the text} used in~\cite{Goseva-Popstojanova2018,Gegick2010,Behl2014,Das2019,Pereira2019}. The idea is to use frequencies of words to retrieve quantitative metrics such as TF/IDF, BF, DF, IDF, etc.
	\item \textbf{Keyword mining} used in~\cite{Peters2019,Mostafa}. The idea is to mine a list of words related to a prediction class and to use this list for pre-filtering and/or for performing prediction results.
\end{itemize}
\subsection{Validation}

In the literature, several validation approaches have been used to assess the performance of the machine learning based approaches for detecting security relevant bug reports. 
Overall, related works perform manual validation or automatic validation using oracle (k-fold cross-validation, random sampling, etc.). 
\begin{itemize}
    \item \textbf{Manual Validation} requires a security team to validate the prediction results. As a result, this validation method is the most reliable in the sense that it provides strong evidence that the approach does not overfit to the training dataset and can be leveraged in the wild (i.e., in a real-world practitioner setting). Unfortunately, manual validation does not scale (i.e., cannot be applied for hundreds of samples since human resources are hard and expensive to get). 
    \item \textbf{Automatic Validation}, due to its ability to scale, is the most widespread in the literature. Traditional techniques such as k-fold cross-validation or random split validation are often used. They are often however criticized since they may not offer a realistic view of the performance (e.g., due to possible overfitting, dataset size limitations, sample distributions problemes, etc.)
\end{itemize}

Automatic validation is by far the most popular technique used to validate a newly proposed security bug report detection approach. However, it is noteworthy that a strong prerequisite is the availability of ground truth. In other words, the availability of a dataset of labeled bug reports (i.e., a dataset in which bug reports are accurately tagged as security relevant or not) is crucial.


\find{{\bf Limitations identified thanks to this review:} We postulate that performance results and conclusions presented in the literature may be biased by some key experimental aspects:
\begin{itemize}
    \item {\em size and availability of datasets}: As detailed in Table~\ref{tab:fps}, experimental datasets are often small or limited in project scope. Large datasets (such as from Microsoft) are not openly available to the community.
    \item {\em lack of rationale guiding algorithm selection}: we have found that most experimental approaches do not motivate the choice of their algorithm, which may lead to redundant approaches in the literature. 
\end{itemize}
$\Rightarrow$
 In this work, we propose to address these limitations for the community by proposing a large dataset (cf. Section~\ref{sec:datasets}) that can enable reproducible research on the identification of security-relevant bug reports. We also consider typical features used in the literature and typical learning algorithms to build and study experimental baseline variations for ML-based detection of security-relevant bug reports (cf. Section
~\ref{sec:experiments})}

\vspace{-1mm}
\subsection{Related work on ``How far are we'' studies}
\vspace{-1mm}
In recent years, the research community has investigated many research topics to produce an estimate of the performance that can be reached. For example, Liu et al.~\cite{liu2018neural} have investigated the case of neural-machine-translation-based commit message generation, while Lin et al.~\cite{lin2018sentiment} focused on sentiment analysis for software engineering. More recently Lou et al.~\cite{lou2019history} presented a "how far are we" study on the problem of history-driven build failure fixing.



\vspace{-1mm}
\subsection{Related work on Bug reports triaging}
\vspace{-1mm}
Bug triaging represents an essential component of the bug handling process. This explains the fact that through literature, several existing works proposed differents approaches improve the process of bugs triaging~\cite{Gadge2017,Tamrawi2011,Kevic2013,Jain2012,Dedik2016,Wang2014,Mani2019}.

Wu et al.~\cite{Wu2011} proposed bugMiner, a tool that able to estimate bug report trend with trend analysis model based on Weibull distribution. BugMiner also allows completing redundancy check on a new or given bug report.  Whang et al.~\cite{Wang2012} built an automatic approach based on semi-supervised learning to identify bugs components by using historical fixed bugs reports. Another work that performs automating bug triaging in literature is done by Jain et al.~\cite{Jain2012}. They proposed an approach using field-based weighting scheme to perform a machine learning tool for bugs assignment. Kevic et al.~\cite{Kevic2013} proposed a collaborative approach to do automated bug triaging by refining the list of relevant developers able to fix a given bug. They use the information retrieval method to identify the bug reports that are similar to the existing triaged bugs. They then analyze the changes sets associated with each bug report and predict the bug nature. Wang et al. ~\cite{Wang2014} investigate unsupervised learning approach for bug triaging, they proposed an approach based on developers’ activeness scores in component of product to build a prioritized list of developers and then improve supervised bug triage approaches (~\cite{anvik2006should}). Dedık et al.~\cite{Dedik2016} did a comparative study on automating bug triaging using SVM classifier and TF/IDF features vectors. They proposed an overview of requirements and needs for automating bug reports in an industrial context and they did a comparative study of their results in industry vs an open-source software project usually used in research: Firefox. 

Most recent, Goyal et al.~\cite{Goyal2019} propose an empirical study of ensemble-based methods for the classification of new bug reports. They studied five ensemble-based classification techniques used in state of art with several machine learning algorithms and show that using ensemble classification techniques outperform single classifiers.

Mani et al.~\cite{Mani2019} go further by investigating the effectiveness of a deep learning approach for bug triaging. They propose a new bug-report representation using a deep bidirectional recurrent neural network with attention (DBRNN-A). Nnamoko et al.~\cite{Nnamoko2019} propose also a deep learning approach to predict automatically labels of a newly reported bug. They used a combination of neural network classification and a similar one vs. all approach to involve training a single classifier per class.


%% file: dataset.tex
\section{Dataset}
\label{sec:datasets}
Machine Learning approaches rely on datasets. Besides the choice of the features, their performance strongly depends on the quality of the dataset used to train and test the obtained model.
After quickly presenting the datasets used in the literature, we will present our own dataset that we named \datasetName.
We propose \datasetName to overcome the limitations of the existing datasets. 
To the best of our knowledge, \datasetName is currently the most comprehensive and largest available dataset for the community. 

\subsection{Datasets in the Literature of Security Bug Report Identification}
As revealed in summary details of Table~\ref{tab:fps}, datasets leveraged by the state of the art approaches are generally limited to a small number of repositories, resulting in a limited number (and under-diversified set) of collected security-relevant bug reports. Most of the identified works in the literature use less than six sources (projects) of bug reports for their experiments. There are even a few works~\cite{wijayasekara2014vulnerability,Behl2014,Pereira2019} that use a single source to predict security-relevant bug reports. Most literature approaches have been applied on datasets containing less than 1\,500 security-relevant bug reports. Zou et al.~\cite{Zou2018} exceptionally included 3\,346 security-relevant bug reports for their work. Pereira et al.~\cite{Pereira2019} have collected a huge {\em title set} of 552\,073 security-relevant bug reports. Unfortunately, the released dataset is incomplete (and thus not reliable for most approaches) since it does not include full description of the bug reports.

With the aforementioned limitations (i.e., few repositories and the limited number of bug reports), the risk is that the proposed approaches may overfit to a few repositories and may not be able to yield comparable performance with data from another repository.  
Since some authors have already demonstrated in literature~\cite{Goseva-Popstojanova2018} of security bug report identification that the quality of the dataset is closely tied to the prediction performance, we propose to build a large and representative dataset of bugs reports for providing the community a playground to advance research directions.

\subsection{\datasetName: a Security-relevant Bug Report Dataset}

We built a large dataset of bug reports that we named \datasetName. 
The \datasetName dataset contains 
5\,028 security-relevant bug reports and 9\,336 non-security relevant bugs reports. 
To the best of our knowledge, \datasetName is the largest dataset in the literature. 
Morever, \datasetName has been built by collecting bug reports from various sources and by various means: Mozilla and RedHat bug reports that include tags for security relevance or not, bug reports linked to commits that were collected and manually assessed by previous works as being related to security issues~\cite{ponta2019manually}, and security bug reports from a literature dataset~\cite{mostafaautomatic}. 

Table~\ref{tab:dataset} summarizes the statistics on the different collection sources.

\begin{table}[!h]
	\centering
	\scriptsize
	\caption{Dataset Details.}
	\resizebox{1\linewidth}{!}{
	\begin{threeparttable}
		\begin{tabular}{p{45mm}p{12mm}p{13mm}r}
		 {\bf Source} & {\bf Security bug reports}& {\bf Non-security bug reports} & {\bf total} \\
		\toprule
		manually collected from 79 Github projects (starting from commits collected by Ponta et al~\cite{ponta2019manually}) &124  &257 & 381\\
		\midrule
		Bug reports available in the dataset proposed by from Mostafa et al~\cite{mostafaautomatic} &  & 9079 & 9079\\
		\midrule
		Mozilla Bugzilla &  202& &202\\
		\midrule
		Redhat &  4702&  & 4702\\
		
		\bottomrule
		All&{\bf 5028}&{\bf 9336}&{\bf 14364}\\
		
		\bottomrule
		\end{tabular}
	\end{threeparttable}
	}
	\label{tab:dataset}
\end{table}



Concretely, a number of security-relevant bug reports have been manually collected by relying on a publicly available security-commits dataset release in 2018~\cite{ponta2019manually}.
The main difficulty in this first part was to link commits to bugs reports. To perform this task, we first parsed and used regular expressions to extract bug ids from the security-relevant commits. Then, we used the API rest call for JIRA-projects (or manual process for other projects), in order to extract bug reports corresponding to the identified bug ids. Note that the dataset released in ~\cite{ponta2019manually} allows us to also collect non-security commits. By following a similar process, these non-security commits have been used to retrieve non-security bug reports.  At the end of this first step, we identified 124 security bug reports and 257 non-security bug reports within 79 Github-hosted projects. 


By checking the bug tracking systems of Mozilla and RedHat projects we were able to rely on the publicly-visible security tags put by development teams to collect 202 and 4\,702 security-relevant bug reports, respectively from Mozilla and RedHat.

A large set of non-security bug reports are obtained from a publicly available dataset published by~\cite{mostafaautomatic}. This dataset includes 9\,079 labeled non-vulnerable bugs reports.

Finally, we standardized the data formats (after curating unreadable bug reports due to encoding issues and dropping incomplete records) and merged all the collected bug reports to obtain labeled dataset that contains 5028 security bugs reports and 9336 non-security bugs reports. Besides the number of security and non-security bug reports, another asset of the \datasetName dataset is that it covers over 80 repositories (Redhat, Mozilla and 79 Github projects). Figure~\ref{fig:br} shows an example of a bug report record in our dataset, which we publicly released at:
\begin{center}
    \url{https://github.com/AAA-create/dataset}.
\end{center}

\begin{figure}[!htb]
	{\parbox{\linewidth}{
\begin{lstlisting}[linewidth={\linewidth},frame=tb,basicstyle=\scriptsize\ttfamily]
{
    "id": "JENKINS-47593",
    "title": "can not to be online without permission",
    "description": "After upgrade to 2.86 plugin need extra 
    permission for dynamic script \u00a0to make slave online
    via ssh connection.\u00a0\r\n\r\nI can not use\u00a0\
    "In-process Script Approval\" because this script look 
    like\u00a0\r\n'ssh user@123.212.12.232 ....' and it is 
    dynamic!\u00a0"
  },
\end{lstlisting}
	}}%
	\caption{Example of a Bug report on our json format dataset}
	\label{fig:br}
\end{figure}

%% file: engineering.tex

\section{Experimental Assessment of ML-based Identification of Security-relevant Bug Reports}
\label{sec:experiments}

The goal of this work is to investigate the performance that can be achieved by reproducing the essential components of literature approaches. The objective is thus to evaluate the performance of various combinations of features (BF, DF, IDF, TF-IDF, etc.) and machine learning algorithms in order to identify what are the key ingredients leading to an efficient security-relevant bug report detector. We select to perform the reproduction of literature ideas (instead of replication of approaches) since our objective is not to compare the state of the art approaches among themselves.

Concretely, we perform a comprehensive study by leveraging the newly created \datasetName dataset (with both security and non-security bug reports), and the identified features and algorithms from literature works. In the following, we first detail the methodology of our study and then present the results. 

\begin{table*}[!t]
\caption{Classification performance results with different combinations of algorithms and feature sets: the full content (title+description) of the bug report is considered as input}
\centering
\resizebox{1.0\linewidth}{!}{%
\begin{tabular}{c|c|c|c|c|c|c|c|c|c|c|c|c}
 &\multicolumn{4} {|c|}{\bf BF}&\multicolumn{4}{|c|}{\bf TF}&\multicolumn{4}{|c}{\bf TF-IDF} \\
 \toprule
 \bf Learners &\bf Accuracy & \bf Precision & \bf Recall&\bf F-score&\bf Accuracy & \bf Precision & \bf Recall&\bf F-score&\bf Accuracy & \bf Precision & \bf Recall&\bf F-score\\
\toprule
XGB & 0.688308&0.544481&0.882427&0.673172 &0.839412&0.809743&0.708201&0.755371&0.979434&0.994105&0.946844&0.969864\\
\rowcolor{grey}
SVC&0.418101&0.384834&0.998138&0.555323 &0.350655&0.350655&1.000000&0.519057&0.881345&0.990082&0.668059&0.797654\\
Decision Tree (DT)&0.687938&0.544249&0.881078&0.672588&0.798199&0.705701&0.727345&0.716188&0.970233&0.962746&0.951859&0.957213\\
\rowcolor{grey}Random Forest (RF)&0.688308&0.544481&0.882427&0.673172&0.847444&0.845274&0.691343&0.760493&0.982014&0.987868&0.960472&0.973944\\
Gaussian Naive Bayes (GNB)&0.546475&0.444359&0.978562&0.610997&0.635025&0.194023&0.012520&0.023493&0.860290&0.802827&0.797215&0.799780\\
\rowcolor{grey}K-Nearest-Neighbors (KNN)&0.401242&0.378142&0.998138&0.548297&0.759920&0.618666&0.821085&0.705445&0.919760&0.879974&0.892604&0.886049\\
Bagged Decision Trees 1 (BDT1)&0.687716&0.544144&0.879497&0.672038&0.844929&0.834244&0.695179&0.758347&0.974136&0.978457&0.947065&0.962426\\
\rowcolor{grey}Bagged Decision Trees 2 (BDT2)&0.687716&0.544144&0.879497&0.672038&0.844929&0.834244&0.695179&0.758347&0.974136&0.978457&0.947065&0.962426\\
Extra Trees (ET)&0.687938&0.544249&0.881078&0.672588&0.847444&0.846640&0.689651&0.760045&0.982502&0.988701&0.961064&\textbf{0.974663}\\
\rowcolor{grey}AdaBoost 1 (ADA1)&0.688751&0.544825&0.884555&0.674039&0.820063&0.759188&0.712291&0.734750&0.978597&0.982016&0.956444&0.969037\\
AdaBoost 2  (ADA2)&0.688751&0.544825&0.884555&0.674039&0.820063&0.759188&0.712291&0.734750&0.978597&0.982016&0.956444&0.969037\\
Gradient Boosting (GB)&0.688308&0.544481&0.882427&0.673172&0.839342&0.811433&0.705528&0.754581&0.976715&0.991426&0.941649&0.965843\\

\bottomrule
\end{tabular}
}
\label{tab:titledesc results}
\end{table*}

\subsection{Methodology}
We propose to perform feature extraction following the approaches proposed in the literature of machine learning for text content. Most of these features are the same as those used by the state of the art approaches for security bug report identification. Then, we propose to focus our investigations on supervised learning, which is the most common model learning scenario in the literature.  We finally attempt a deep learning experiment as an initial result that the community can build on.

\subsubsection{Feature extraction}
In this work, we build four types of features vectors: Binary Bag-of-words (BF), Term Frequency (TF), Term Frequency-Inverse Document Frequency (TF-IDF) and Word Embedding vectors (WE). For each bug report, we then compute these features according to their respectively equations.
\begin{itemize}
    
\item \textbf{Binary Bag-of-Words (BF)} is defined by the following equation:
\begin{equation}
BF(term)=\left\{
\begin{array}{l}
  0, if f(term)=0 \\
  1, if  f(term)>0
\end{array}
\right.
\end{equation}
with \textit{BF(term)} being the binary bag-of-word of a term in a given document and f(\textit{term}) the frequency of this term. Simply, for each given bug report, Binary bag-of-word determines the presence or not of words of the collected vocabulary (from all bug reports).
\item \textbf{Term Frequency (TF)} is defined by the following equation:
\begin{equation}
TF(term) = f(term)
\end{equation}
TF gives the frequency of a word apparition for each document. Words are enumerated from a given vocabulary (e.g., the set of all words appearing across all bug reports).

\item \textbf{Term Frequency-Inverse Document Frequency (TF-IDF)} is defined by the following equation:
\begin{equation}
TF-IDF(\textit{term})=f(term)* log \frac{|D|}{N(term)}
\end{equation}
with $N(term)$ being the number of documents where  $term$  appears and $|D|$ the total number of documents in the corpus. In other words, Term Frequency-Inverse Document Frequency (TF-IDF) is a measure of how much information the word provides, that is, whether it is common or rare across all bug reports.

\item\textbf{Word embedding} is common in natural language processing approaches. It allows to represent words as numerical vectors so that similar words are close in the vector space. We relied on the default embedding implementor of Tensorflow\footnote{tensorflow.feature.text} to compute the word embeddings of bug reports.
\end{itemize}



In our methodology, we focus only on the textual description of bug reports. No other artefacts (e.g., which developer was assigned to the bug report) to avoid leaking information in the training. Indeed, we consider the identification of security-relevant bug reports as a zero-day problem: the only information available is the report that was submitted before anyone from the development team could add insightful comments or data. 

Since we are performing {\em early detection of security bug reports}, we extract three contents that are available at the time the bug report is submitted: {\tt Title} content, {\tt Description} content, and the concatenation of these two contents as {\tt Title+Descrition} content. For each content, we compute TF, TF-IDF, BF and WE vectors. 
Note that before building the feature vectors, some preprocessing operators have been applied: we removed stop words (e.g., “as”, “is”, “would”, etc.) which are very frequently appearing in any English document and will not hold any discriminative power; we also do stemming (only consider the root of words) using Porter Stemming~\cite{willett2006porter} method. 


\subsubsection{Model learning}
Our experiments are focused on supervised learning with manual feature engineering or deep representation learning for feature vector inference.

\textbf{Supervised learning}: We experiment with eleven (11) different supervised algorithms on our features sets. These are Decision Tree, Decision Trees (Bagged), Random Forest, Extra Trees, Adaptive Boosting, Gradient Boosting, eXtreme Gradient Boosting, Gaussian Naive Bayes, Support Vector Classifier, K-Nearest-Neighbors, and Stochastic Gradient Descent. All these classifiers are used with their default hyperparameter values from the {\em scikit-learn} library, and we compute the performance metrics based on 5-fold cross-validation (given to the datase size).

\begin{figure}[!h]
\vspace{-4mm}
\centering
\includegraphics[width=0.70\linewidth]{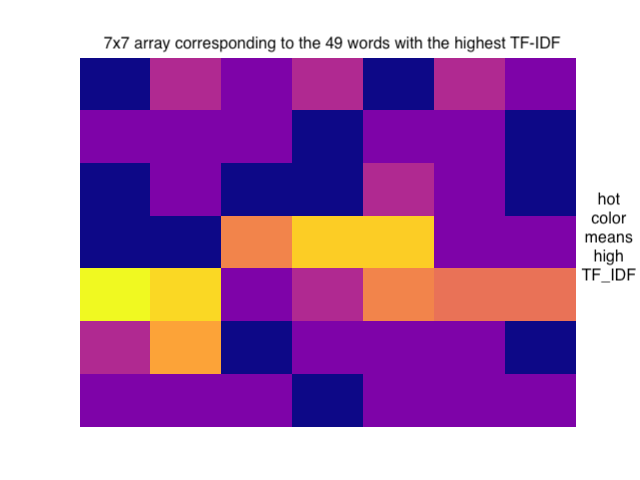}
\vspace{-5mm}
\caption{Heatmap representation of a bug report (with hight TF-IDF values being represented with hot colors - towards yellow)}
\label{fig:CNN}
\vspace{-3mm}
\end{figure}

\textbf{Deep learning}: The use of deep learning in security detection is not much studied in the state of the art. We propose however to experiment a straightforward approach by leveraging  Long Short Term Memory (LSTM)~\cite{hochreiter1997long} and Convolutional Neural Network (CNN)~\cite{kalchbrenner2014convolutional}. Since we already use word embedding for feature extraction, we can apply LSTM to the classification problem of security relevance. Nevertheless, we resort to applying CNN on the best data format that it has been shown to be successful on, namely images. Thus, we use Text2Image\footnote{https://towardsdatascience.com/text2image-a-new-way-to-nlp-cbf63376aa0d} method to convert bug reports to images for training and testing the CNN model. 
More specifically,  after a pre-processing step (tokenization, stop-word removal, etc.), Text2Image builds 7x7 arrays by considering the 49 words with the highest TF-IDF. We note here that the number of words chosen can be considered as a hyper-parameter. Finally, the images are plotted as heatmaps on log scale using Gaussian filtering for smoothness.  Figure
~\ref{fig:CNN} shows an example of a security bug report heatmap.  The hottest the color is,  the highest the TF-IDF value is. Given such heatmaps, we use CNN to train a deep learning classifier to discriminate security-relevant bug reports from other bug reports. 


\input{results.tex}

%% file: results.tex
\begin{table*}[!t]
\caption{Classification  performance  results  - considering only bug report \underline {title} as input}
\centering
\resizebox{1.0\linewidth}{!}{%
\begin{tabular}{c|c|c|c|c|c|c|c|c|c|c|c|c}
 &\multicolumn{4} {|c|}{\bf BF}&\multicolumn{4}{|c|}{\bf TF}&\multicolumn{4}{|c|}{\bf TF-IDF} \\
 \toprule
 &\bf Accuracy &\bf  Precision &\bf  Recall&\bf F-score&\bf Accuracy & \bf Precision & \bf Recall&\bf F-score&\bf Accuracy & \bf Precision & \bf Recall&\bf F-score\\
\rowcolor{grey}XGB &  646831&0.000000&0.000000&0.000000&0.646760&0.050000&0.000202&0.000402&0.836797&0.926826&0.579682&0.713186\\
SVC &0.355277&0.353896&0.999802&0.522596&0.353169&0.353169&1.000000&0.521831&0.824108&0.921228&0.544363&0.684287\\
\rowcolor{grey}Decision Tree (DT) &0.646831&0.000000&0.000000&0.000000&0.648868&0.513572&0.108945&0.179563&0.838888&0.935031&0.580317&0.716079\\
Random Forest (RF)&0.646831&0.000000&0.000000&0.000000&0.648939&0.513650&0.109938&0.180896&0.838888&0.933670&0.581335&0.716449\\
\rowcolor{grey}Gaussian Naive Bayes (GNB)&0.374386&0.360492&0.996714&0.529329&0.513488&0.407471&0.830950&0.546570&0.756344&0.696359&0.539417&0.607770\\
K-Nearest-Neighbors (KNN)&0.357595&0.354690&0.999411&0.523402&0.403541&0.368955&0.969869&0.534385&0.487035&0.403159&0.963674&0.568305\\
\rowcolor{grey}Bagged Decision Trees 1 (BDT1)&0.646831&0.000000&0.000000&0.000000&0.648728&0.512229&0.109938&0.180803&0.839028&0.934505&0.581139&0.716550\\
Bagged Decision Trees 2 (BDT2)&0.646831&0.000000&0.000000&0.000000&0.648728&0.512229&0.109938&0.180803&0.839028&0.934505&0.581139&0.716550\\
\rowcolor{grey}Extra Trees (ET)& 0.646831&0.000000&0.000000&0.000000&0.649712&0.519581&0.108945&0.179926&0.838958&0.934527&0.580919&0.716388\\
AdaBoost 1 (ADA1)&0.646831&0.000000&0.000000&0.000000&0.650273&0.574718&0.088977&0.149848&0.835124&0.918708&0.580569&0.711449\\
\rowcolor{grey}AdaBoost 2  (ADA2)&0.646831&0.000000&0.000000&0.000000&0.650273&0.574718&0.088977&0.149848&0.835124&0.918708&0.580569&0.711449\\
Gradient Boosting (GB)&0.646831&0.000000&0.000000&0.000000&0.646550&0.490137&0.037963&0.064262&0.838888&0.931980&0.582545&\textbf{0.716901}\\

\bottomrule
\end{tabular}
}
\label{tab:title results}
\end{table*}

\vspace{-2mm}
\begin{table*}[!t]
\caption{Classification  performance  results  - considering only bug report \underline {description} as input}
\centering
\resizebox{1.0\linewidth}{!}{%
\begin{tabular}{c|c|c|c|c|c|c|c|c|c|c|c|c}
 &\multicolumn{4} {|c|}{\bf BF}&\multicolumn{4}{|c|}{\bf TF}&\multicolumn{4}{|c}{\bf TF-IDF} \\
\toprule
 \bf Learners &\bf Accuracy & \bf Precision &\bf Recall& \bf F-score& \bf Accuracy & \bf Precision &\bf  Recall&\bf F-score&\bf Accuracy & \bf Precision & \bf Recall&\bf F-score\\
 \toprule
\rowcolor{grey}XGB & 0.628535&0.523802&0.817072&0.638091&0.842338&0.833315&0.742255&0.785039&0.975530&0.989473&0.940135&0.964129 \\
SVC &0.480961&0.435743&0.995330&0.605892&0.388573&0.388573&1.000000&0.559479&0.923174&0.976550&0.799684&0.879195\\
\rowcolor{grey}Decision Tree (DT) &0.628535&0.523802&0.817072&0.638091&0.810247&0.744537&0.778833&0.761145&0.966118&0.952127&0.951117&0.951570\\
Random Forest (RF)&0.628535&0.523802&0.817072&0.638091&0.853267&0.851863&0.753385&0.799333&0.979434&0.981087&0.959850&0.970315\\
\rowcolor{grey}Gaussian Naive Bayes (GNB)&0.614021&0.509875&0.968319&0.667838&0.711738&0.514806&0.373145&0.396211&0.754181&0.609551&0.831535&0.703242\\
K-Nearest-Neighbors (KNN)&0.431061&0.413578&0.996777&0.584274&0.810324&0.707844&0.871196&0.780832&0.872348&0.769005&0.929133&0.838228\\
\rowcolor{grey}Bagged Decision Trees 1 (BDT1)&0.628535&0.523802&0.817072&0.638091&0.848384&0.841239&0.751776&0.793717&0.974415&0.976273&0.950151&0.962969\\
Bagged Decision Trees 2 (BDT2)&0.628535&0.523802&0.817072&0.638091&0.848384&0.841239&0.751776&0.793717&0.980410&0.985714&0.958066&\textbf{0.971652}\\
\rowcolor{grey}Extra Trees (ET)& 0.628535&0.523802&0.817072&0.638091&0.855360&0.861863&0.747174&0.800260&0.980410&0.985714&0.958066&\textbf{0.971652}\\
AdaBoost 1 (ADA1)&0.628535&0.523802&0.817072&0.638091&0.824510&0.810038&0.716594&0.759808&0.976018&0.979926&0.950854&0.965151\\
\rowcolor{grey}AdaBoost 2  (ADA2)&0.628535&0.523802&0.817072&0.638091&0.824510&0.810038&0.716594&0.759808&0.976018&0.979926&0.950854&0.965151\\
Gradient Boosting (GB)&0.628535&0.523802&0.817072&0.638091&0.844353&0.836646&0.744321&0.787696&0.972532&0.984765&0.936025&0.959744\\

\bottomrule
\end{tabular}
}
\label{tab:description results}
\end{table*}

\subsection{Overall Results}
We present the empirical results for the different classifiers that we have built. Besides the feature extraction methods and the algorithms, we have also varied the feature vector sizes as well as the target bug report content (e.g., title or description) to assess the impact of the different variations.

\begin{center}
    {\bf Exploring features and algorithms}
\end{center}
Table~\ref{tab:titledesc results} shows the classification performance results that are obtained for each pairwise combination of learning algorithms and feature sets. We note that the F-score performance has a wide range between 0.67 and 0.97 depending on the combination of feature set and algorithm. TF-IDF based feature set offers the best performance  (between 0.79 and 0.97). Aside from Naive Bayes (GNB), Nearest Neighbors (k-NN), and Support Vector Classifier (SVC), all other algorithms  (generally tree-based) perform similarly with high classification F-scores. 

\find{The choice of feature set has the biggest impact on the classification performance of security-relevant bug reports. TF-IDF offering the best results. Tree-based learners are the best for security-relevant classification, and yield similar performance results.}

\vspace{-3mm}
\begin{center}
    {\bf Exploring variations of inputs from bug reports}
\end{center}
\vspace{-2mm}
Our review of the state of the art has revealed that some approaches claim to yield the best results when focusing on part of the bug report (such as the title). We propose to redo all experiments combining different algorithms with different feature engineering targets, however focusing on changing the input content to be considered from bug reports. 
Table \ref{tab:title results} shows the results yielded when using only the bug report title to classify security-relevant bug reports. 
TF-IDF is still revealed as the best method for engineering features for classifying security bug reports, and can help to achieve over 0.71 F-score with tree-based learners. These results confirm the recent findings made by Pereira et al.~\cite{Pereira2019} on \underline{closed Microsoft data}: it is possible to achieve reasonably good prediction performance by only leveraging bug report titles. The best F-score performance (0.71), using titles, however, is 26 percentage points lower than the best performance (0.97) obtained with full bug report contents.

\find{Although, in our experiments, classification with full contents of bug reports leads to the best results, bug report titles appear to carry sufficient information for yielding reasonable classification performance.}

Similarly, we have undertaken to assess whether report titles are redundant within a bug report. Thus, we perform the same experiments as above considering only the bug report descriptions as input. Table~\ref{tab:description results} summarizes the obtained performance results. TF-IDF based feature set is further confirmed as being the most adapted. Note that in this case, however, the best performance is only at less than 1 percentage point from the best performance obtained with the full bug report content. 

\find{Experimental results show that bug report description contains most of the information, which if captured with TF-IDF based features, will lead to accurate classification of security relevance.}

\vspace{-2mm}
\begin{center}
    {\bf Exploring deep learned features}
\end{center}
\vspace{-1mm}
Table~\ref{deeplearning:lstm} provides accuracy results of using LSTM approaches with word embeddings to predict security bug reports. Training loss and validation loss at the different epochs are sensibly similar, allowing us to discard immediate reasons that suggest overfitting or underfitting. The obtained \textbf{accuracy} is generally high (although generally slightly lower than the ones obtained based on engineered features in previous experiments). This suggests that Word embedding and LSTM are a good combination to accurately detect security bug reports. Accuracy is indeed between  87\% and 90\%.

\begin{table}[hhh]
\caption{Classification results with LSTM + word embeddings}
\centering
\resizebox{0.6\linewidth}{!}{%
\begin{tabular}{c|c|c|c}
\bf Epoch  &\bf Training loss&\bf Validation loss& \bf Accuracy\\
\toprule
\rowcolor{grey}0  & 0.2855&0.0000&0.8714\\
6 & 0.2905&0.2877&0.8739\\
\rowcolor{grey}10 & 0.2696&0.3102&0.8774\\
16  & 0.2387&0.2380&0.8853\\
\rowcolor{grey}20  & 0.1981&0.2040&\textbf{0.9062}\\

\bottomrule
\end{tabular}
}
\label{deeplearning:lstm}
\end{table}

\vspace{-2mm}
Results of our alternate deep learning approach that leverages TF-IDF heatmaps with CNN to extract structural image features for classification are provided in 
Table~\ref{deeplearning:cnn}. The accuracy is slightly higher than the precedent LSTM-based approach, ranging between  88\% and 92\%.

\begin{table}[hhhh]
\caption{Classification results with CNN + TF-IDF heatmaps}
\centering
\resizebox{0.6\linewidth}{!}{%
\begin{tabular}{c|c|c|c}
\bf Epoch  & \bf Training loss & \bf Validation loss & \bf Accuracy\\
\toprule
\rowcolor{grey}0  & 0.219164 & 0.308599 & 0.904665\\
1 & 0.254183 & 0.346557 & 0.888438\\
\rowcolor{grey}2 & 0.259294&0.296789&\textbf{0.926978}\\
3  & 0.210394&0.300965&0.922921\\
\rowcolor{grey}4  & 0.157864&0.302180&0.922921\\

\bottomrule
\end{tabular}
}
\label{deeplearning:cnn}
\end{table}

\vspace{-10mm}
\find{Deep representation learning supported approaches can achieve high performance, irrespective of the type of inputs (tokens or heatmaps) that are fed to the neural nets.}

\vspace{-2mm}
\begin{center}
    {\bf Exploring Feature vector sizes}
\end{center}

We propose to investigate the importance of the vector-size parameter in security bug reports identification based on engineered features. 
We experimentally focus on traditional supervised learning algorithms studied earlier in this section using the TF-IDF based feature set, which was shown to provide the best results.

\begin{figure*}[!h]
\centering
\includegraphics[width=0.90\linewidth]{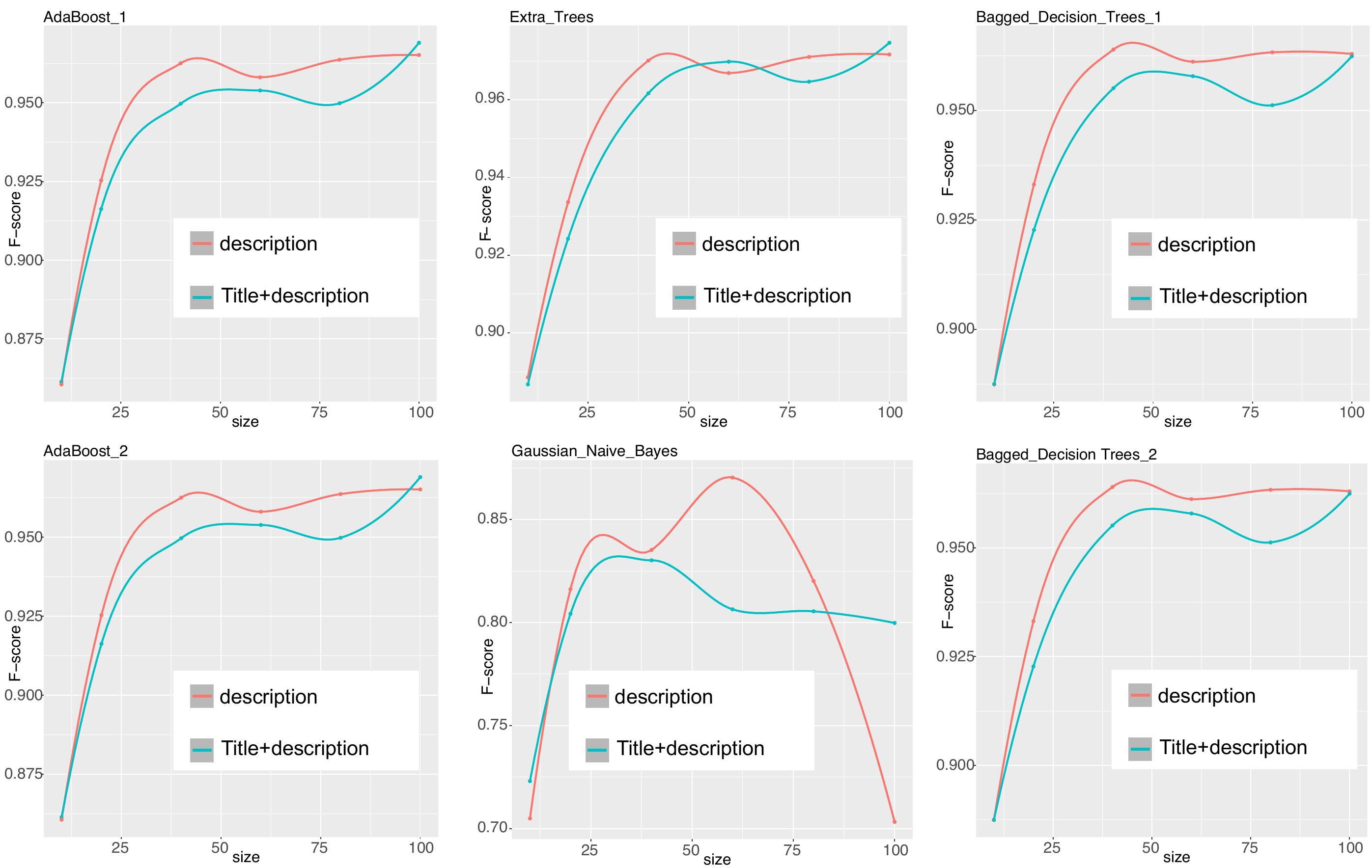}
\caption{Results when the feature-vector size increases}
\label{fig:vectorSize}
\end{figure*}

Figure~\ref{fig:vectorSize} shows the evolution of F-score values as a function of TF-IDF feature vector size for the top learners (all are tree-based except Gaussian Naive Bayes). 
As inputs, we considered the description content (the orange curve) and the full content (``tile + description'') (blue curve).

The experimental data reveal that the feature vector size has a direct impact on the F-score values: the performance generally increases with the feature vector size, but appears to cap at some point (e.g., beyond a size of 50-80, the performance is somewhat stabilized). We even observe that there is a window of feature size for which the performance can decrease again before going back to the plateau, suggesting that the amount of noisy features must be controlled when trying to reduce the feature vector size.

Interestingly, we note that the  performance of classifiers that are based on Gaussian Naive Bayes, can be severely impacted by large feature vectors. This result may explain the limited performance yielded in the comparative experiments by non-tree based classifiers: these learners are less capable to model security-relevance when too many features are considered. 


\find{{\bf feature-vector size}: The prediction F-measure evolves with the size of feature vectors.}

%% file: related.tex
\section{Discussion}
\label{sec:discussion}
In this section, we first discuss the threats to validity involved in our work. 
We then summarize the take-home messages. 
Finally, we discuss related studies as well as related approaches in the real of bug report triaging.

\vspace{-1mm}   
\subsection{Threats to Validity}
\vspace{-1mm}
Our work carries some threats to validity that we have tried to mitigate.

In terms of threats to {\bf external validity}, we note that the \textbf{dataset may be noisy}. Although, we have assembled  datasets from the literature, some of  them may be unreliable. Indeed, for example, Ponta et al~\cite{ponta2019manually} have identified security-relevant bug reports to build their dataset by following a manual process. We reduced this threat by double-checking their manual collection process. Similarly, our negative set is collected from the literature. Unfortunately, the heuristics of collection by authors are not always available for cross-checking.

In terms of threats to {\bf internal validity}, we note that our  selection of learning algorithms may be biased by the trends in the literature. We have mitigated this threat by considering 11 supervised learning algorithms, different neural architectures as well as different input formats.

We also have some threats to {\bf construct validity} related to the implementation of deep learning approaches.
Hyper-parameters such as epochs, the size of most discriminant words, etc. create biases in the model. To reduce this bias, we performed extensive tests for hyper-parameter selection and reported training and validation losses at different epochs. 

\vspace{-1mm}
\subsection{Take-Home Messages Summary}
\vspace{-1mm}
 


    
Our empirical evaluation on our collected dataset shows that:
\begin{itemize}
\item Features based on TF-IDF yield the best classification results. 
\item Tree-based classifiers perform generally well for the classification of security-relevant bug reports with TF-IDF feature set.
\item The performance of non-tree based classifiers can be affected by a large and noisy feature set. 
\item As suggested by previous experiments on proprietary data, we confirm on large open source data that bug report title includes information to enable prediction of security relevance. However, most relevant information tokens are in the description.  
\end{itemize}

We also offer initial results on the use of straightforward deep learning models for the classification of security-relevant bug reports. Our ambition is to allow researchers to build on these insights to build approaches that can be integrated to development environments.

\find{Our experimental results suggest that, {\bf in the lab}, machine learning-based classifiers can accurately predict security-relevant bug reports. The challenge now for the research community is to experimentally demonstrate this achievement {\bf in the wild}, by integrating and assessing such classifiers in a production bug triaging pipeline.}

%% file: conclusion.tex
\section{Conclusion}
\label{sec:conclusion}
Quickly addressing security bug reports is paramount. Indeed, when users report bugs, these reports are addressed with varying depending on the tags that developers give them. 
An unnoticed security bug report may further cause damages if they are exploited by attackers before the development teams can handle them. 
In this work, we propose an overview of the state of art in security bug reports identification. We dissect the most recurrent components and show the achievements made in the literature on a large dataset.
We underlined some factors (mainly feature engineering) that can lead to the best identification of security bug reports. 
As future work, we plan to integrate  security-relevant bug reports classifiers in open source repositories, and assess these classifiers in production bug triaging pipeline. 